# The 2D Spin and its Resonance Fringe

B. C. Sanctuary

*Department of Chemistry*
*McGill University*
*Montreal, Quebec H3A 2K6*
*Canada*

**Abstract.** Violation of Bell's Inequalities gives experimental evidence for the existence of a spin ½ which has two simultaneous axes of spin quantization rather than one. These couple to form a resonance state, called the spin fringe, and this quantum effect is solely responsible for violation of Bell's Inequalities within this model. The Bell states can be represented by products of these spin states and leads to the intuitive concept that as entangled states decompose they form biparticles that are not entangled. In EPR coincidence experiments filter settings for both the Bell and CHSH forms of Bell's Inequalities are rationalized in terms of the correlation between biparticles.

## I. INTRODUCTION

Recently it has been shown that the four Bell states are separable in terms of products of non-hermitian operators that describe the states of single spins rather than ensembles [1]. At that level all a spin's attributes are simultaneously dispersion free and deterministic [2]. However in an isotropic environment, each of the two eigenvalues of ±1 is eight-fold degenerate. Therefore it is not possible to know, or likely worth knowing, which particular state a spin occupies, leading to a possible origin of the statistical nature which is the level that quantum theory (QT) describes.

In this paper the states of the two dimensional (2D) spin model are discussed further and it is shown that ensembles of particles filtered in EPR coincidence experiments [3] not only reproduce the results of QT but also can rationalize those filter settings that lead to violations of Bell's Inequalities (BI) [4],[5]. It is also shown that coherence exists between the two orthogonal axes of spin quantization. This is manifest as a resonance, or quantum exchange term, which has magnitude √2 and corresponds to a spin angular momentum of magnitude $\hbar/\sqrt{2}$. It is called the *spin fringe* because it results from a single spin interfering with itself [6]. Finally it is shown that phase randomization destroys the spin fringe and leads to hermitian state operators that always obey BI. Therefore we conclude that non-hermiticity is the fundamental cause of the violation of BI. Before proceeding, the results of a previous paper [1] are summarized in order to provide the background for the treatment here.

Rather than assuming that the Bell states persist to infinite separation, it is more intuitive to expect correlated spin pairs, called biparticles, are formed from some process of separation which is depicted by,

$$\rho_{\text{Bell}} \xrightarrow{\text{A single decomposition}} \rho^1_{\hat{\mathbf{n}}} \rho^2_{\hat{\mathbf{n}}} \qquad (1.1)$$

That is a Bell state decomposes into product states formed from two spins. Therefore the state operators at this level do not represent ensembles but individual particles. The subscript $\hat{\mathbf{n}}$ denotes a Cartesian coordinate system, $\hat{\mathbf{n}} = \{\hat{\mathbf{x}}, \hat{\mathbf{y}}, \hat{\mathbf{z}}\}$, which is called the spin *microframe* within which a single spin is defined. Biparticles are formed with identical microframes, and in the absence of external interactions, the microframes of each biparticle maintain their common orientation as they separate. The correlation between biparticles is therefore created at the source and is maintained up until they are filtered. In EPR coincidence experiments [3] the filters at Alice and Bob are assumed to obey Malus's law and are considered to be sufficiently separated that one filter has no influence on the other [2].

A single spin has a well defined microstate at any instant and this is described by the non-hermitian state operator,

$$\rho_{\hat{n}}(n_z, n_x, n_y) \equiv \frac{1}{2}(I + n_z \sigma_{\hat{z}} + n_x \sigma_{\hat{x}} + n_y i \sigma_{\hat{y}}) \tag{1.2}$$

which is motivated by recognizing that the diagonal elements of a singlet state are projections,

$$\rho_{\hat{z}}(n_z) = \frac{1}{2}(I + n_z \sigma_z) \tag{1.3}$$

and the off-diagonal elements are coherences, $\sigma_x \pm i\sigma_y$. These are combined to define Eq.(1.2). The parameters introduced in addition to $n_z$ are local hidden variables, (LHV), $n_x, n_y$, and the set $n_{zxy} \equiv (n_z, n_x, n_y) = \{\pm 1, \pm 1, \pm 1\}$ has eight permutations corresponding to the octants of the microframe. In each octant the eigenvalues are always $\pm 1$ and there are always two simultaneous eigenvectors, one associated with the $\hat{z}$ axis and the other with the $\hat{x}$ axis: for example in the first octant,

$$\boldsymbol{\sigma} \cdot [\hat{z} + \hat{x} + i\hat{y}] = \begin{pmatrix} 1 & 2 \\ 0 & -1 \end{pmatrix} \text{ gives } +1 \text{ with } \begin{pmatrix} 1 \\ 0 \end{pmatrix} \text{ and } -1 \text{ with } \frac{1}{\sqrt{2}} \begin{pmatrix} 1 \\ -1 \end{pmatrix} \tag{1.4}$$

The non-hermiticity is essential for these non-orthogonal eigenvectors and these states, along with the recognition that $\sigma_z$ and $\sigma_x$ are components of angular momentum, lead to the conclusion that the intrinsic spin of an electron has two orthogonal axes of spin quantization and therefore is two dimensional, see Figure 1(a). The non-orthogonality of the states in Eq.(1.4) also suggests that a single electron can self-interfere.

In the above treatment there are no states associated with the term causing the non-hermiticity, $i\sigma_y$. This is a rotation operator and not a component of angular momentum in this model. It orients the 2D spin in three dimensional space and it is called the *quantum operator phase*, or simply the quantum phase. It is not possible to directly detect an orientation but it can be obtained indirectly from the commutation relation, $\sigma_z \sigma_x = i\sigma_y$. Therefore rather than the usual set of spin attributes of $\sigma_x, \sigma_y, \sigma_z$, here they are taken as $\sigma_x, i\sigma_y, \sigma_z$.

Equation (1.2) can be written as,

$$\rho_{\hat{n}}(n_{zxy}) = \frac{1}{2}(I + Q_{cl} \boldsymbol{\sigma} \cdot \hat{\mathbf{n}}_{zxy}) \tag{1.5}$$

where the unit vector is defined,

$$\hat{\mathbf{n}}_{zxy} = \frac{1}{\sqrt{3}}(n_z \hat{z} + n_x \hat{x} + i n_y \hat{y}) \tag{1.6}$$

The parameter, $Q_{cl}$, is called the *quantum correlation length*, (QCL) and in Eq.(1.5), it is $\sqrt{3}$. It is simply the normalization factor of the unit vector $\hat{\mathbf{n}}_{zxy}$. A vector of length $\sqrt{3}$ in the direction of $\hat{\mathbf{n}}_{zxy}$ has unit projections on the microframe axes, and trisects the microframe and is therefore called the *microframe axis*. By varying the local variables $n_z, n_x, n_y$, $\hat{\mathbf{n}}_{zxy}$ spans the octants.

Biparticle states are formed from taking products of single spin states, Eq.(1.1). By varying the local variables over the octants of such products, the four Bell states can be represented. For example the singlet Bell states is separable as a sum of eight biparticle states [1],

$$\rho_{\Psi_{12}^-} = \frac{1}{8} \begin{bmatrix} \rho_{\hat{n}}^1(+1,+1,+1)\rho_{\hat{n}}^2(-1,-1,+1) + \rho_{\hat{n}}^1(+1,+1,-1)\rho_{\hat{n}}^2(-1,-1,-1) + \\ \rho_{\hat{n}}^1(+1,-1,+1)\rho_{\hat{n}}^2(-1,+1,+1) + \rho_{\hat{n}}^1(+1,-1,-1)\rho_{\hat{n}}^2(-1,+1,-1) + \\ \rho_{\hat{n}}^1(-1,+1,+1)\rho_{\hat{n}}^2(+1,-1,+1) + \rho_{\hat{n}}^1(-1,+1,-1)\rho_{\hat{n}}^2(+1,-1,-1) + \\ \rho_{\hat{n}}^1(-1,-1,+1)\rho_{\hat{n}}^2(+1,+1,+1) + \rho_{\hat{n}}^1(-1,-1,-1)\rho_{\hat{n}}^2(+1,+1,-1) \end{bmatrix} \quad (1.7)$$

where the Bell state operator is defined,

$$\rho_{\Psi_{12}^-} = |\psi_{12}^-\rangle\langle\psi_{12}^-| = \frac{1}{2}(I^1 I^2 - \boldsymbol{\sigma}^1 \cdot \boldsymbol{\sigma}^2) \quad (1.8)$$

in terms of the usual singlet,

$$|\psi_{12}^-\rangle = \frac{1}{\sqrt{2}}\left(|+\rangle_z^1|-\rangle_z^2 - |-\rangle_z^1|+\rangle_z^2\right) \quad (1.9)$$

The other three Bell states can be similarly represented [1].

In summary, this approach shows that the states of a single spin are deterministic because the expectation values of the set $\sigma_x, i\sigma_y, \sigma_z$ are dispersion free using the state operators for spin microstates, Eq.(1.2). At the level that ensembles are formed however, the eight-fold degeneracy in this spin model due to the LHVs give a rational for the statistical nature of QT. Moreover the results of calculating the correlation between EPR pairs using either the product representation Eq.(1.7) or the entangled singlet Eq.(1.8) are equivalent so that the product representation agrees with the predictions of QT. From Eq.(1.7) all biparticles that represent the Bell states have identical microframes. In EPR coincidence experiments, [3] biparticles are formed at the source and then separate into beams thereby forming ensembles. These are filtered into sub-ensembles, and as shown in this paper, the 2D spin model is consistent with filter settings that give maximum violation of BI. Violations of BI are shown to occur only when the spin fringe is present. If it is randomized, BI are always satisfied.

## 2. THE SPIN FRINGE

The 2D spin is defined in one of the eight equivalent octants and has states associated with two simultaneous axes of quantization: one associated with $\sigma_x$ and the other with $\sigma_z$ see Ea.(1.4). These two spin axes are indistinguishable, see Figure 1(a), thereby creating a $C_2$ symmetry axis about the direction that bisects the two spin components. This leads to an exchange or resonance term which is identified as a single spin resonance fringe or simply, the spin fringe (superscript *SF* in Eq.(2.1)). This superposed state is formed by two states with well defined but opposite quantum phases which give a hermitian state,

$$\rho_{\hat{n}}^{SF}(n_{zx}) = \frac{1}{2}\left(\rho_{\hat{n}}(n_{zxy}) + \rho_{\hat{n}}(n_{zx-y})\right) = \frac{1}{2}\left(I + \sqrt{2}\boldsymbol{\sigma}\cdot\hat{\mathbf{n}}_{zx}\right) \quad (2.1)$$

The unit vector in each quadrant formed from by the two spin axes $\hat{\mathbf{z}}$ and $\hat{\mathbf{x}}$ is,

$$\hat{\mathbf{n}}_{zx} = \frac{1}{\sqrt{2}}(n_z\hat{\mathbf{z}} + n_x\hat{\mathbf{x}}) \quad (2.2)$$

where now the QCL is reduced from √3 to √2. In contrast to the coherent states, there are only two local variables that define the quadrants, $(n_{zx}) \equiv (n_z, n_x) = \{\pm 1, \pm 1\}$, wherein a spin lies along with its spin fringe, but the parameter,

$n_y$, although not explicitly shown, is required to have a definite value in order to produce the fringe as seen from Eq.(2.1). Resonance states are common in the quantum mechanical treatment of, for example, chemical bonding [7] where the exchange term makes a significant contribution to molecular stability. Here this effect is responsible to the formation of an additional angular momentum that, as shown below, is the cause of violation of BI in this model.

From Eq.(2.1), the four sets of states, $|\pm\rangle^{SF}_{n_z n_x}$, are given by:

Two sets of orthogonal states for $\pm\sigma_{\hat{z}} \pm \sigma_{\hat{x}}$

$$\pm\frac{1}{\sqrt{2}}\begin{pmatrix} 1 & 1 \\ 1 & -1 \end{pmatrix} \text{ gives eigenvalues of } \pm 1 \qquad |\pm\rangle^{SF}_{\pm 1,\pm 1} = \frac{1}{\sqrt{2}\sqrt{\left(1+\frac{1}{\sqrt{2}}\right)}}\begin{pmatrix} 1+\frac{1}{\sqrt{2}} \\ \frac{1}{\sqrt{2}} \end{pmatrix} \qquad (2.3)$$

with

$$\pm\frac{1}{\sqrt{2}}\begin{pmatrix} 1 & 1 \\ 1 & -1 \end{pmatrix} \text{ gives eigenvalues } \mp 1 \qquad |\mp\rangle^{SF}_{\pm 1,\pm 1} = \frac{1}{\sqrt{2}\sqrt{\left(1+\frac{1}{\sqrt{2}}\right)}}\begin{pmatrix} \frac{1}{\sqrt{2}} \\ -1-\frac{1}{\sqrt{2}} \end{pmatrix} \qquad (2.4)$$

and two sets of orthogonal states for $\pm\sigma_{\hat{z}} \mp \sigma_{\hat{x}}$

$$\pm\frac{1}{\sqrt{2}}\begin{pmatrix} 1 & -1 \\ -1 & -1 \end{pmatrix} \text{ gives eigenvalues of } \pm 1 \qquad |\pm\rangle^{SF}_{\pm,1\mp,1} = \frac{1}{\sqrt{2}\sqrt{\left(1+\frac{1}{\sqrt{2}}\right)}}\begin{pmatrix} -1-\frac{1}{\sqrt{2}} \\ \frac{1}{\sqrt{2}} \end{pmatrix} \qquad (2.5)$$

with

$$\pm\frac{1}{\sqrt{2}}\begin{pmatrix} 1 & -1 \\ -1 & -1 \end{pmatrix} \text{ gives eigenvalues } \mp 1 \qquad |\mp\rangle^{SF}_{\pm,1\mp,1} = \frac{1}{\sqrt{2}\sqrt{\left(1+\frac{1}{\sqrt{2}}\right)}}\begin{pmatrix} \frac{1}{\sqrt{2}} \\ 1+\frac{1}{\sqrt{2}} \end{pmatrix} \qquad (2.6)$$

Geometrically these eigenstates lie at angles of 45° between the four sets of two spin axes defined by $n_z\hat{z}$ and $n_x\hat{x}$. The spin fringe lies in the quadrants wherein the two axes of quantization lie. These two spin components superpose to create a spin fringe of length √2. Therefore a spin angular momentum is defined in units of $\hbar/2$ as

$$\mathbf{S}_{\sqrt{2}} \equiv \frac{\hbar}{2}\sqrt{2}\boldsymbol{\sigma} = \frac{\hbar}{\sqrt{2}}\boldsymbol{\sigma} \qquad (2.7)$$

revealing an irrational quantum number. This spin has only one dispersion free state for each of the four vectors where $\hat{\mathbf{n}}_{zx}$ defines its axis of quantization,

$$\langle \boldsymbol{\sigma}\cdot\hat{\mathbf{n}}_{zx}\rangle = \text{Tr}\left[\rho^{SF}_{\hat{\mathbf{n}}}(n_{zx})\boldsymbol{\sigma}\cdot\hat{\mathbf{n}}_{zx}\right] = \pm\sqrt{2} \qquad (2.8)$$

Except for its magnitude, the spin density operator for the spin fringe is similar to the usual spin and is one dimensional since it has only a single axis of spin quantization, see Eq.(2.2). Due to hermiticity of Eq.(2.1) these eigenstates are orthogonal in each of the four directions of $\hat{\mathbf{n}}_{zx}$, and are non-orthogonal to the spin states of which Eqs.(1.4) is an example. Figure 1(a) shows the spin components in one octant of a microframe which also displays the spin fringe of length √2 and the microframe axis of length √3. Figure 2 gives the geometry of the 2D spin.

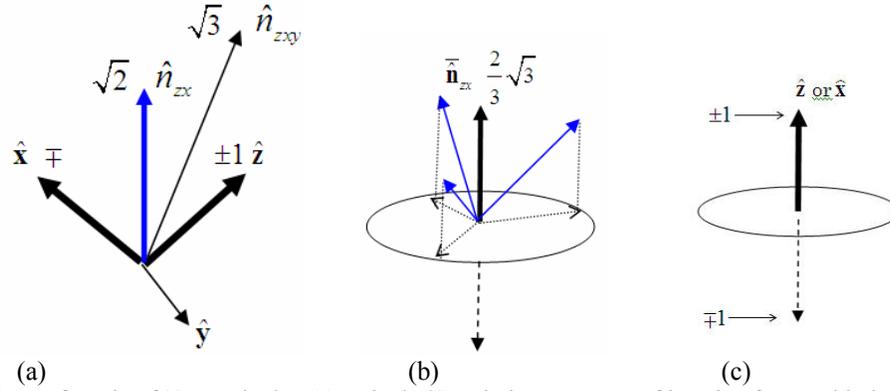

**Figure 1.** Three views of a spin of ½ magnitude. **(a)** a single 2D spin in one octant of its microframe with the microframe axis shown with QCL of √3, and the spin fringe with QCL of √2. **(b)** An ensemble of phase randomized spins quantized along the microframe axis with QCL of (2/3)√3. **(c)** An ensemble of spins with one axis of spin quantization formed from averaging the other axis of quantization and randomizing the phase. Figure (c) is the usual view of a spin as prepared for measurement by a polarizing field.

The $C_2$ symmetry of the two dimensional spin also admits the difference,

$$\frac{1}{2}\left(\rho_{\hat{n}}\left(n_{zxy}\right) - \rho_{\hat{n}}\left(n_{zx-y}\right)\right) = in_y \sigma_y = \pm i\sigma_y \tag{2.9}$$

This is recognized as the quantum operator phase which has no handle for interaction and as such cannot be directly measured. Its two values of ±1 correspond to the two indistinguishable orientations of the 2D spin, which differs by a rotation of ±π.

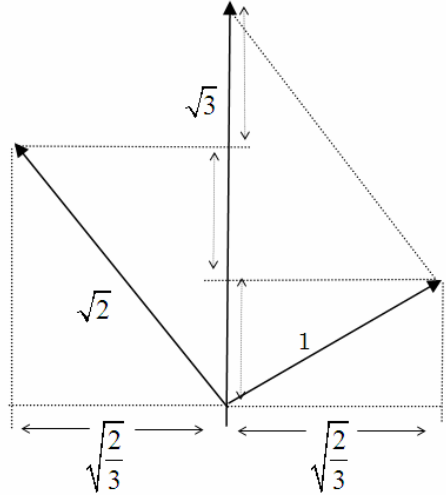

**Figure 2.** Shown is a slice through the two dimensional spin. The component of length 1 is one of the three coordinates of the microframe along which the quantum phase lies. The vertical component is the microframe axis. The spin fringe lies along the component of length √2. As the QCL drops from √3 to √2 to 1, the projection on the microframe axis drops by one third.

## 3. PHASE RANDOMIZATION

For microstates of a spin ½ the maximum QCL is √3 but other cases, such as phase randomization and averaging, lead to reduced values of the QCL as fewer states are resolved. Phase randomization removes the non-hermiticity from the spin microstates and forms an ensemble of hermitian spin states. The existence of the spin fringe requires the quantum phase to be fixed, and it vanishes when the phase is randomized. This is now discussed and constitutes a statistical treatment of a single spin microstate.

The phase, being quantized, can only take three positions between the three coordinate axes of a spin's microframe, see Figure 1(b). The ensemble has only three members in each octant,

$$\rho_{\hat{\mathbf{n}}}^{a}(n_{zxy}) \equiv \frac{1}{2}\left(I + \boldsymbol{\sigma} \cdot (n_z\hat{\mathbf{z}} + n_x\hat{\mathbf{x}} + in_y\hat{\mathbf{y}})\right)$$
$$\rho_{\hat{\mathbf{n}}}^{b}(n_{zxy}) \equiv \frac{1}{2}\left(I + \boldsymbol{\sigma} \cdot (n_z\hat{\mathbf{y}} + n_x\hat{\mathbf{z}} + in_y\hat{\mathbf{x}})\right) \quad (3.1)$$
$$\rho_{\hat{\mathbf{n}}}^{c}(n_{zxy}) \equiv \frac{1}{2}\left(I + \boldsymbol{\sigma} \cdot (n_z\hat{\mathbf{x}} + n_x\hat{\mathbf{y}} + in_y\hat{\mathbf{z}})\right)$$

The usual definition is used for an ensemble [8] which is a collection of non-interacting replicas of the system. Following the methods of statistical mechanics each member of the ensemble is evaluated separately and then the results from each member are averaged over the ensemble. In this case, each member is assumed equally probable and the expectation value for an operator $A$ in a phase randomized microstate is therefore given by,

$$\overline{A}^{1D} = \frac{1}{3}\left(\text{Tr}\left[A_{zx}^{a}\rho_{\hat{\mathbf{n}}}^{a}(n_{zxy})\right] + \text{Tr}\left[A_{yz}^{b}\rho_{\hat{\mathbf{n}}}^{b}(n_{zxy})\right] + \text{Tr}\left[A_{xy}^{c}\rho_{\hat{\mathbf{n}}}^{c}(n_{zxy})\right]\right) \quad (3.2)$$

The superscript $1D$ indicates phase randomization and is considered a one dimensional average because it randomizes one local variable, $n_y$. The state operator now does not represent a single particle but an ensemble of three particles all with the same microframe but with random phases. Each member has only two measurable quantities [9], or some algebra of them. For an electron the two orthogonal quantization axes have equal and opposite magnetic moments, $\mu$, and measurable quantities are $A_{zx}^{a} = \mu\sigma_z$ and $-\mu\sigma_x$; $A_{yz}^{b} = -\mu\sigma_y$ and $\mu\sigma_z$; $A_{xy}^{c} = \mu\sigma_x$ and $-\mu\sigma_y$; $B_{xy}^{c} = -\mu^2\sigma_x\sigma_y = -i\mu^2\sigma_z$, *etc*. Once again, the quantum phase cannot be directly measured. Working out the expectation values for those attributes with a magnetic moment using Eq.(3.2) gives,

$$\overline{\langle\sigma_z\rangle}^{1D} = \overline{\langle\sigma_x\rangle}^{1D} = \overline{\langle\sigma_y\rangle}^{1D} = +\frac{1}{3}(n_z + n_x) = 0, \pm\frac{2}{3} \quad (3.3)$$

These values do not depend on the local variable $n_y$. Because of this, phase randomization introduces dispersion. For the zero expectation value, the dispersion is the maximum at $\pm 1$ while for the other two it is $\pm\sqrt{5}/3 \approx \pm 0.75$. Of the possible states available to a single isolated spin, phase randomization removes them all as elements of physical reality. Their projection, see Figure 1(b), lies along the microfame axis with reduced quantum correlation length of $\sqrt{2}$,

$$\hat{\mathbf{n}}_{zxy} = \overline{\hat{\mathbf{n}}}_{zx} = \frac{1}{\sqrt{2}}\overline{(n_z\hat{\mathbf{z}} + n_x\hat{\mathbf{x}})}^{1D} \quad (3.4)$$

That is, phase randomization obscures the quantum phase; reduces the number of states by one third; and randomizes the spin fringe. There remains only one dispersion free state in the direction of the microframe axis. Therefore the spin state so produced is given by

$$\rho_{\hat{\mathbf{n}}}(n_{zx}) \equiv \overline{\rho_{\hat{\mathbf{n}}}(n_{zxy})}^{1D} = \frac{1}{2}\left(I + \sqrt{2}\boldsymbol{\sigma} \cdot \hat{\mathbf{n}}_{zxy}\right) \quad (3.5)$$

It is important to distinguish this from Eq.(2.1) which is the resonance state for the spin fringe in the direction that bisects the spin along the direction of the unit vector $\hat{\mathbf{n}}_{zx}$, while Eq.(3.5) is quantized along $\hat{\mathbf{n}}_{zxy}$ and results from phase randomization. The former case has three simultaneous dispersion free spin attributes, $\sigma_x, i\sigma_y, \sigma_z$, and the latter has only one, $\sigma_{\hat{\mathbf{n}}_{zxy}}$.

In order to obtain the usual one dimensional spin states, here called *macrostates*, observed by invasive filtering, for example by the use of quarter wave plates or a Stern-Gerlach magnetic field, a second averaging occurs that removes, or obscures, one of the two components of angular momentum as an element of objective reality. For electrons in a magnetic field, one spin component aligns and is dispersion free, while the other, being perpendicular, precesses and averages to zero (see the appendix). This state is described by the state operator discussed by Fano [10] in 1957 which is obtained here from a two dimensional average over the pure microstates and is identical to the projection operator of Eq.(1.3).

$$\rho_z(n_z) \equiv \overline{\rho_{\hat{n}}(n_{zxy})}^{2D} = \frac{1}{2}(I + n_z Q_{cl} \boldsymbol{\sigma} \cdot \mathbf{z}) \qquad Q_{cl} = 1 \qquad (3.6)$$

The 2D averages removes the two local variables, $n_x$ and $n_y$. This operator describes the usual spin with two states; has two eigenvalues of +1 or -1; has no hidden variables; and represents pure states with $Q_{cl}$ of unity. This case is depicted in Figure 1(c). Clearly mixed states correspond to a $Q_{cl} < 1$.

## 4. EVIDENCE FOR THE 2D SPIN AND ITS QUANTUM FRINGE

After Bell derived [4] his inequalities experiments were devised and performed to show quantum predictions to be correct and his inequalities violated. This has been well reviewed [11], [12]. Violations of BI are interpreted here to mean the spin has two rather than one axis of quantization. As mentioned in the introduction, the treatment here assumes that entangled states are produced at the source in EPR coincidence experiments and separate into products of biparticles which are not entangled. If these remain undisturbed from the source to the filters then the two spins that form a biparticle maintain coincident microframes. On the other hand along the way phase randomization or other disturbances are possible and these destroy correlation between biparticles. In this section the correlation is calculated in different situations for the singlet state and these are compared with the experimental results; in particular with those filter settings that lead to violation of BI.

The well known result from QT [4] that assumes the singlet state, Eq.(1.8), persists to space-like separation gives,

$$E(\mathbf{a},\mathbf{b}) = \langle \sigma_\mathbf{a}^1 \sigma_\mathbf{b}^2 \rangle = \mathbf{a} \cdot \langle \boldsymbol{\sigma}^1 \boldsymbol{\sigma}^2 \rangle \cdot \mathbf{b} = \mathbf{a} \cdot \mathrm{Tr}_{12}\left[\boldsymbol{\sigma}^1 \boldsymbol{\sigma}^2 \rho_{\Psi_{12}^-}\right] \cdot \mathbf{b} = -\mathbf{a} \cdot \mathbf{b} = -\cos\theta_{ab} \qquad (4.1)$$

where the filters are set in the directions of **a** and **b** and are assumed not to affect the state operators, [2]. By virtue of Eq.(1.7) the same result is obtained using separable biparticles showing that the 2D spin model agrees with the predictions of QT.

Inspection of Eq. (1.7) reveals that for each biparticle the quantum phase is the same but the two axes are always opposite from one octant to the other. It is therefore possible to define a biparticle in one octant to calculate the correlation between a specific pair and then, assuming spatial isotropy, ensemble average the result. Such a biparticle product state is defined by

$$\rho_{12}^P(n_{zxy}) = \frac{1}{2}\left(\rho_{+\hat{n}}^1 \rho_{-\hat{n}}^{2\dagger} + \rho_{-\hat{n}}^1 \rho_{+\hat{n}}^{2\dagger}\right) = \frac{1}{4}\left(I^1 I^2 - Q_{cl}^2 \boldsymbol{\sigma}^1 \cdot \hat{\mathbf{n}}_{zxy} \hat{\mathbf{n}}_{zxy}^* \cdot \boldsymbol{\sigma}^1\right) \qquad (4.2)$$

where † represents the adjoint operation and * complex conjugation. Ensemble averaging leads to,

$$\overline{\rho_{12}^P(n_{zxy})} = \frac{1}{2}\overline{\left(\rho_{+\hat{n}}^1 \rho_{-\hat{n}}^{2\dagger} + \rho_{-\hat{n}}^1 \rho_{+\hat{n}}^{2\dagger}\right)} = \frac{1}{4}\left(I^1 I^2 - Q_{cl}^2 \boldsymbol{\sigma}^1 \cdot \overline{\hat{\mathbf{n}}_{zxy} \hat{\mathbf{n}}_{zxy}^*} \cdot \boldsymbol{\sigma}^1\right)$$

$$= \frac{1}{4}\left(I^1 I^2 - \frac{Q_{cl}^2}{3} \boldsymbol{\sigma}^1 \cdot \boldsymbol{\sigma}^1\right) = \rho_{\psi_{12}^-} \qquad (4.3)$$

showing that the ensemble averaged product state composed of microstates is indistinguishable from the singlet, Eq.(1.8) for $Q_{cl} = \sqrt{3}$. The correlation is given by

$$E(\mathbf{a},\mathbf{b}) = -\frac{Q_{cl}^2}{3}\cos\theta_{ab} \tag{4.4}$$

If the biparticles undergo phase randomization then the QCL is reduced to √2, Eq.(3.5), and the correlation in Eq.(4.2) is reduced by one third. If the usual spin state is used with $Q_{cl}=1$, Eq.(3.6), the correlation is reduced by two thirds. This last case also results when the entanglement in the singlet is ignored [13]. Table 1 summarizes the results.

**Table (1)** The correlation is evaluated using Eq.(4.4) with different quantum correlation lengths. The last line gives the correlation from the singlet Bell state, Eq.(4.1). A random distribution of spins is assumed for the ensemble average for the first three cases.

| State | Correlation | Elements of reality | $\frac{Q_{cl}^2}{3}$ | Bell's Inequalities satisfied |
|---|---|---|---|---|
| Prepared by Probe | $-\frac{1}{3}\cos\theta_{ab}$ | 1 | One third | yes |
| Phase-randomized | $-\frac{2}{3}\cos\theta_{ab}$ | 1 | Two thirds | Yes exactly |
| Microstate | $-\cos\theta_{ab}$ | 3 | 100% | no |
| Singlet State | $-\cos\theta_{ab}$ | 0 | 0 | n/a |

From these entries, BI are only violated when no averaging or randomization occurs. We therefore conclude that the spin fringe, or equivalently the non-hermiticity of the state, is responsible for the violation of BI.

## Strong Correlation Between Microstates

Bell's Inequalities relate correlation between three events in Bell's original form [4] and four events in the CHSH form [14]. In this section BI are used to study the correlation between biparticles formed from the singlet state. As one photon is filtered at Alice, its bipartner is filtered at Bob thereby forming a measurement co-plane. When the filters are set along directions where the projected spin components lie, enhanced coincidence counting is observed. Specifically settings, say $\mathbf{a},\mathbf{b}$, filter a statistically large number of biparticles which are coincidence counted and averaged to obtain the correlation $E(\mathbf{a},\mathbf{b})$. That is, over the course of an experiment ensembles of biparticles are formed and filtering picks out a specific sub-ensemble for detection. The experiments are repeated with filter settings of $\mathbf{a},\mathbf{c}$ and $\mathbf{b},\mathbf{c}$. In this section, using BI, various filter settings are shown to be consistent with the predictions from correlated 2D spins. The following well known example makes this clear.

One set of filter settings in Bell's Inequalities corresponding to the conditions for strong violation [11] are $\mathbf{a}$ and $\mathbf{c}$ making an angle of 120° and $\mathbf{b}$ making an angle of 60° between each of $\mathbf{a}$ and $\mathbf{c}$. These lead to a violation of,

$$\left|E(\mathbf{a},\mathbf{b}) - E(\mathbf{a},\mathbf{c})\right| \leq 1 + E(\mathbf{b},\mathbf{c}) \quad \text{or} \quad \frac{Q_{cl}^2}{3}\left|-\frac{1}{2}-\frac{1}{2}\right| \lessgtr 1 - \frac{1}{2}\frac{Q_{cl}^2}{3} \quad \text{or} \quad 2 \lessgtr 1 \text{ for } Q_{cl}^2 = 3 \tag{4.5}$$

Strong correlation occurs in this case when the microframe axis is perpendicular to the measurement co-planes, see Figures 3a and 3b. The projection of the two axes of quantization, $\hat{\mathbf{z}}$ and $\hat{\mathbf{x}}$ onto the co-plane gives them 120° apart and of projected length $(2/3)^{1/2}$, see Figure 2. If one spin is filtered along direction $\mathbf{a}$ corresponding to the projected part of, say $\hat{\mathbf{z}}$ for this sub-ensemble, then its bipartner will have its component $\hat{\mathbf{x}}$ projected 120° from $\mathbf{a}$ in its co-plane, thereby giving strong correlation if the filter setting $\mathbf{c}$ is 120° from $\mathbf{a}$ see Figure 3. This corresponds to one event in BI, Eq.(4.5).

In addition, the spin fringe is directed along $\hat{\mathbf{n}}_{zx}$ bisecting the vectors $\hat{\mathbf{z}}$ and $\hat{\mathbf{x}}$ with QCL of √2, see Figure 3(b). This component projects onto the measurement co-planes making an angle of 60° between $\mathbf{a}$ and $\mathbf{c}$ with the same projected length of $(2/3)^{1/2}$, see Figure 2. Strong correlation is therefore expected for filter settings that differ by

±60° thus being consistent with this particular violation of Bell's Inequalities, Eq.(4.5). In addition, this gives experimental support for the existence of the spin fringe, (see Figure 3(b)). Since the quantum phase, $i\sigma_y$, cannot be measured, it is not projected onto the measurement co-plane.

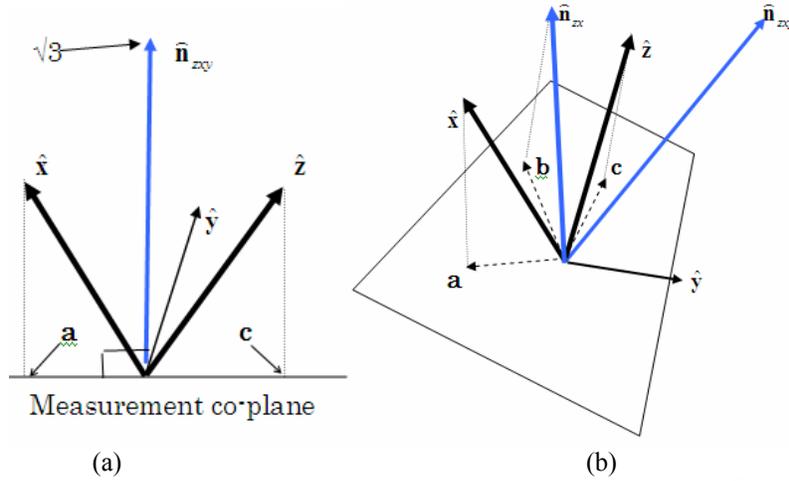

(a)                          (b)

**FIGURE (3).** (a) Shows strong correlation between two particles occurs when a sub-ensemble of particles all with microframe axes perpendicular to the measurement co-planes with projections of $z$ and $x$ being 120° apart. The phase component is not projected onto the co-plane because an orientation cannot be filtered. The spin resonance is suppressed for clarity. (b) is the same as (a) but in perspective. Now the spin resonance has been added. The projection of the spin fringe onto the measurement co-plane is shown by **b** which bisects **a** and **c**.

## Suppression of the Spin Fringe

In addition to the strong correlation found above for the 60° settings, other filter settings are sensitive to the correlation between biparticles. First, see Figure 1(a), correlation exists along the direction of $\hat{\mathbf{n}}_{zx}$ where the spin fringe lies. If the two dimensional spin has its axis $\hat{\mathbf{n}}_{zx}$ perpendicular to the measurement co-plane rather than $\hat{\mathbf{n}}_{zxy}$, then Figure 4 is obtained. There the components of each spin pair, $\hat{\mathbf{z}}$ and $\hat{\mathbf{x}}$, are projected 180° apart with length of $(\sqrt{2})^{-1}$ while $\hat{\mathbf{n}}_{zx}$ has projected length of zero and bisects $\hat{\mathbf{z}}$ and $\hat{\mathbf{x}}$. The spin fringe is therefore suppressed and cannot be detected at this setting. Hence filter settings of **a** and **c** 180° apart and **b** making an angle of 90° between each of **a** and **c** lead to $\mathbf{a} \cdot \mathbf{c} = -1$ and, not surprisingly since $\hat{\mathbf{n}}_{zx}$ has zero projection, $\mathbf{a} \cdot \mathbf{b} = \mathbf{b} \cdot \mathbf{c} = 0$. These setting are not sensitive to the spin fringe and so Bell's Inequalities are not violated but are exactly satisfied. These angles give,

$$\begin{aligned}|E(\mathbf{a},\mathbf{b}) - E(\mathbf{a},\mathbf{c})| &\leq 1 + E(\mathbf{b},\mathbf{c}) \\ |-0-1| &= 1+(0)\end{aligned} \quad (4.6)$$

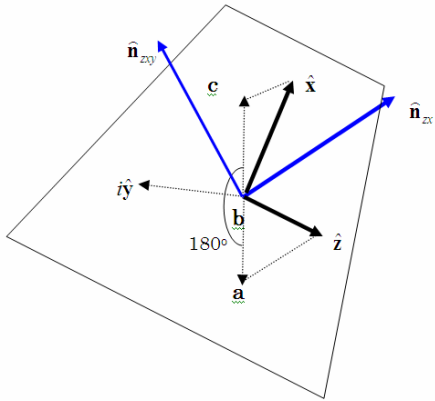

**Figure (4).** Correlation between microstates along the direction of the spin fringe. Now the quantum operator phase is parallel to the detection co-plane. Since the spin fringe axis, $\hat{\mathbf{n}}_{zx}$, is perpendicular to the measurement co-plane, the correlation due to the spin fringe is suppressed and BI are not violated but exactly satisfied.

The spin fringe can also be suppressed by phase randomization. Recall the quantum operator phase, $i\sigma_y$, is quantized, and when it is randomized it can take only three positions. Phase randomization can possibly occur due to some process between the source and before filtering. Figure 5 depicts the situation discussed in Eq.(3.1) which illustrates three filter settings of 120° would emerge in this case. However because of the destruction of the spin fringe, BI cannot be violated. Phase randomization reduces correlation between the biparticles by one third as seen from Figure 5 and Table (1). This leaves only a one dimensional hermitian state of angular momentum with $Q_{cl} = \sqrt{2}$, Eq.(3.5). Since the correlation for the phase randomization case is reduced by one third, the three filter angles differing by 60° exactly satisfy BI,

$$|E(\mathbf{a},\mathbf{b}) - E(\mathbf{a},\mathbf{c})| \leq 1 + E(\mathbf{b},\mathbf{c})$$
$$\frac{Q_{cl}^2}{3}\left|-\frac{1}{2} - \frac{1}{2}\right| = 1 + \frac{Q_{cl}^2}{3}\left(-\frac{1}{2}\right) \quad (4.7)$$
$$\frac{2}{3} = 1 - \frac{1}{3}$$

This is the same as Eq.(4.5) except that the QCL has been reduced to √2 for the phase randomized case.

The discussions in this subsection show that the difference between violation and non-violation of BI is due to the presence of the spin fringe. If the quantum phase is suppressed or randomized, BI are always satisfied for ensemble averaging over a random distribution. Violations of BI therefore give experimental support for the existence of the quantum operator phase, the spin fringe and the deterministic microstate of a single spin.

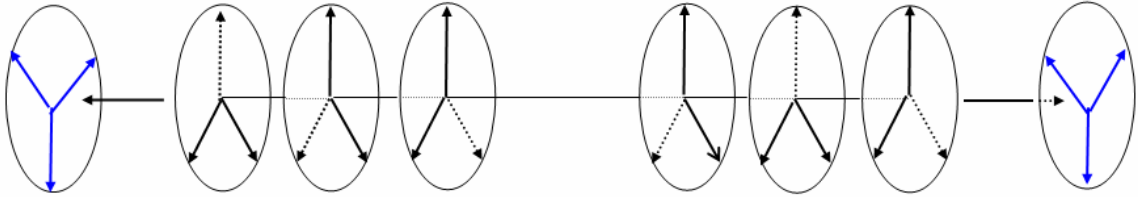

**Figure (5).** Loss of correlation by phase randomization before filtering. In this case the spins retain correlation only between the two spin components and the spin fringes vanish. The correlation is reduced by one third. Bell's Inequalities are always satisfied for filter settings that differ by 120°.

## Violation of the CHSH form of Bell's Inequalities.

The CHSH [14] form of Bell's Inequalities is more general than Bell originally derived because Bell did not include the case of perfect anti-correlation, $E(\mathbf{b},\mathbf{b}) = 1$ Including this point leads to the CHSH form,

$$|E(\mathbf{a},\mathbf{b}) - E(\mathbf{a},\mathbf{c}) + E(\mathbf{b},\mathbf{b}') + E(\mathbf{b}',\mathbf{c})| \leq 2 \quad (4.8)$$

The maximum violation occurs with four filter settings of 45° apart and contiguous,

$$\left|-\frac{1}{\sqrt{2}} - \left(+\frac{1}{\sqrt{2}}\right) - \frac{1}{\sqrt{2}} - \frac{1}{\sqrt{2}}\right| = 2\sqrt{2}\,\frac{Q_{cl}^2}{3} \not\leq 2 \quad (4.9)$$

Another way of viewing the correlation between biparticles is when the quantum phase is perpendicular to the measurement co-plane, see Figure 6(a). Hence settings the filters to **a**, **b** and **c** being 45° apart will be sensitive to these three spin components. However the CHSH form of BI introduces a forth setting: **a, b, b′** and **c**. From figure 6(b) if the difference between **b** and **b′** is taken to be 45° then the correlation is enhanced due to the overlap of two independent correlations within the statistical system that have a common quantum phase. That is one event is associated with correlation between one pair of biparticles with settings of **a, b** and **b′**. The second pair of biparticles, independent from the first but with the same quantum phase, are detected at settings of **b, b′** and **c**, thereby overlapping and enhancing the visibilities of the **b**, **b′** settings. Since the two sets of biparticles must have a common quantum phase, their correlations depends on the existence of microstates with $Q_{cl} = \sqrt{3}$.

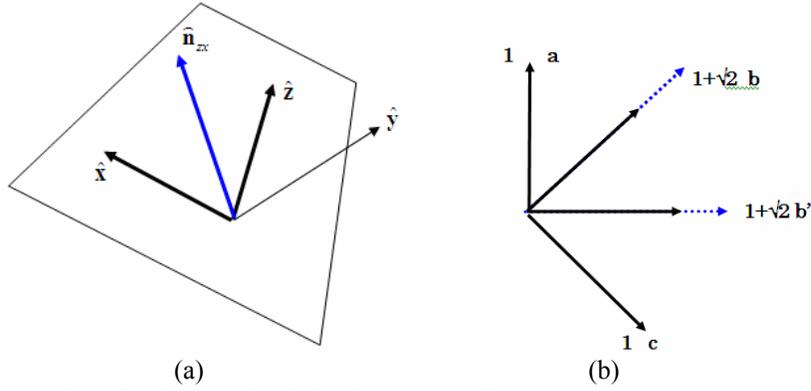

(a) (b)

**FIGURE 6.(a)** A sub-ensemble of spins all with their phases perpendicular to the measurement co-plane so the two spin components and its resonance are all parallel to the co-plane and set at 45° apart. (**b**) Two independent spin components in the measurement co-plane with one rotated from the second by 45°.

The two filter settings of 60° and 45° cause violations in both the Bell and CHSH forms but to different extents. Collapsing the CHSH settings so that **b** = **b**′ can then be used in Bell's form with **a**, **b** and **c** set 45° apart. Then the violation is

$$|E(\mathbf{a},\mathbf{b}) - E(\mathbf{a},\mathbf{c})| \leq 1 + E(\mathbf{b},\mathbf{c})$$

$$\left| -\frac{1}{\sqrt{2}} \frac{Q_{cl}^2}{3} - 0 \right| \leq 1 - \frac{1}{\sqrt{2}} \frac{Q_{cl}^2}{3} \qquad (4.10)$$

$$2.414 \not\leq 1$$

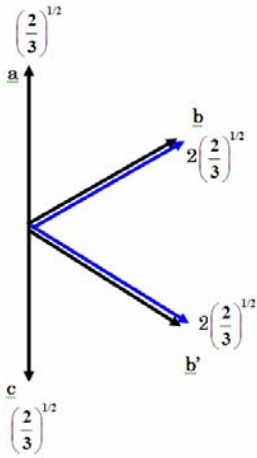

which gives a greater violation than for settings of 60° apart, i.e., $2 \not\leq 1$ see Eq.(4.5).

Similarly it is possible to overlap two independent biparticle correlations by taking the settings of 60° apart and rotating one from the other by 60° degrees rather than 45°, see Figure 7. However the enhancement of this **b**, **b**′ setting is less than for 45° because the projected part for the 60° settings have lengths of only $(2/3)^{1/2}$ which is considerably less that for 45° of $1+\sqrt{2}$, see Figure 6(b). This leads to greater intensities for the 45° settings relative to the 60° ones. Consequently using the 60° settings in the CHSH form gives a violation of $1.25 \not\leq 1$ which is less than for the 45° settings, see Eq.(4.9), of $1.414 \not\leq 1$.

**Figure 7** Two independent sets of coincidence are shown with one differing from the other by 60°. Hence the two overlap thereby enhancing the correlations between the **b** and **b**′ settings, but to a lesser extent than for settings that are 45° apart due to the reduced projected lengths, *cf* Figure (6b).

In summary, the Bell form and the CHSH form of Bell's Inequalities with different filter settings probe different sub-ensembles of the same correlation that exists between biparticles. Strong correlation is expected between biparticle states when the three different axes are perpendicular to the measurement co-plane. These are identified: $\hat{\mathbf{n}}_{zxy}$ with QCL of $\sqrt{3}$; $\hat{\mathbf{n}}_{zx}$ with QCL of $\sqrt{2}$; and the quantum phase $i\sigma_y$ with QCL of 1. It appears no other filter settings will give stronger correlation. Equations (4.5), (4.9) and (4.10) all violate Bell's Inequalities for QCL of $\sqrt{3}$ but all satisfy them for $\sqrt{2}$. violation of BI therefore support the existence of spin microstates and the fringe.

## 5. GEOMETRIC CORROBORATION OF THE SPIN FRINGE

Using noncommutative trigonometry to study the CHSH form of Bell's Inequalities, Gustafson [15] has recently found vectors of magnitude $\sqrt{2}$ which are consistent with the existence of the spin fringe and are responsible for the violation of BI. From the CHSH form he obtains an expression from which he can find conditions that give maximum violation for specific choices of the angles between the detector settings,

$$\left| \mathbf{b}' \cdot (\mathbf{b}+\mathbf{c}) + \mathbf{a} \cdot (\mathbf{b}-\mathbf{c}) \right| = 2\left( \cos^2 \theta_{b',b+c} + \cos^2 \theta_{a,b-c} \right)^{1/2} \left| \cos \theta_{u_1,u_2} \right| \qquad (5.1)$$

He finds the angle between **b** and **c** must be $\pm \pi/2$ and then chooses the other angles such that $\cos^2 \theta_{a,b-c} = \cos^2 \theta_{b',b+c} = 1$. Under these conditions, two vectors emerge such that the region where the stronger-than-classical correlation occurs corresponds to $1 \leq \mathbf{u}_1 \cdot \mathbf{u}_2 \leq \sqrt{2}$. If these two vectors are collinear then maximum violation occurs.

Gustafson's analysis is consistent with the 45° settings that maximize the CHSH form. Figure 8 is the same as Figure 6(b) with the sum and difference $(\mathbf{b} \pm \mathbf{c})$ from Eq. (5.1) included. This figure shows that the angle between **b** and **c** is π/2, consistent with Gustafson's findings and moreover that $(\mathbf{b}+\mathbf{c})$ is parallel to $\mathbf{b}'$ and $(\mathbf{b}-\mathbf{c})$ is parallel to **a**, in agreement with $\cos^2 \theta_{a,b-c} = \cos^2 \theta_{b',b+c} = 1$. These angles in Gustafson's formulae lead to the conclusion the vectors that give maximum violation bisect the quadrants of length √2,

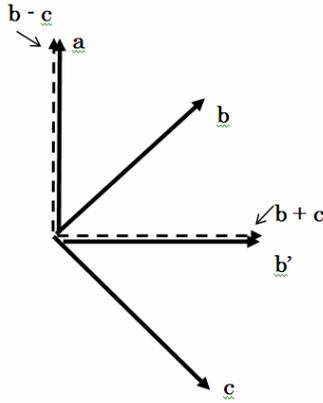

$$\mathbf{u}_1 = \frac{1}{\sqrt{2}}(\pm 1, \pm 1) \qquad (5.2)$$

Without using quantum mechanics, this purely geometric analysis of Bell's Inequalities shows that maximum violation of the CHSH form requires the existence of an extra vector component which is identified here as the spin fringe.

**Figure 8.** Using the 45° angles that maximize the violation of the CHSH form of Bell's inequalities it is found that the sum and difference between the vectors shown are parallel to respectively **a** and $\mathbf{b}'$. Analysis by Gustafson leads to the conclusion that four vectors of length √2 bisect each of the quadrants in agreement with the properties of the two dimensional spin.

## 6. CONCLUSIONS

This paper studies some of the properties of the two dimensional spin of magnitude ½ which contains a spin resonance or interference fringe. This purely quantum term arises due to the indistinguishability of the two orthogonal spin axes of quantization and is an angular momentum with two states of magnitude $\hbar/\sqrt{2}$. The findings here are consistent with EPR coincidence experiments but further experimental work focusing on the spin properties, rather than non-locality [16],[17],[18] is needed to more thoroughly test this model.

The statistical interpretation of quantum mechanics [19],[20] provides a physical basis for rationalizing the experimental results and avoids paradoxes [21]. All spin attributes are simultaneously elements of physical reality [2] for microstates so probability is not needed. However probability is needed in QT [22] and is found here to be a consequence of the lack of resolution of some local variables, see *e.g.* Eq.(3.3), due to effects such as phase randomization and averaging, which occurs naturally or as a result of preparation for direct measurement [19]. In such processes the hidden variables cannot be individually resolved and ensembles are formed. In contrast EPR coincidence experiments [3] are non-invasive so biparticles are not disturbed between the source and the filters. Therefore these experiments are sensitive to the properties of the 2D spin and the correlation between non-entangled biparticles albeit a large number are collected and analyzed statistically. Entanglement is not, therefore, a property of nature but an essential and unique [23] property of quantum mechanics whereby a few states can statistically account for those that cannot be resolved.

Other LHVM have been shown to agree with the predictions of QT [24],[25],[26],[27],[28],[29] but these cannot rationalize the filter settings that lead to violation of BI. All these models appear to introduce LHV which are consistent with the probabilities obtained from QT. However, besides probability being obviated for individual spins, it is impossible to deduce non-hermitian states from models that reproduce the quantum probabilities. The approach in this paper is different because it starts from the non-hermitian states of single particles and applies the methods of statistical mechanics [8] to build up ensembles.

It goes contrary to complementarity [30] that all attributes of a particle are simultaneously elements of physical reality. Until the advent of reliable EPR data [3] the one dimensional view of a spin could not be disputed and there was no evidence that LHV existed. The main purpose of this paper is to show that the EPR experiments are sensitive to LHV and are consistent with the existence of a second axis of spin and the interference fringe. However this requires a fundamental change to the hermitian postulate of quantum mechanics. The 2S spin model indicates that violation of Bell's Inequalities means the spin fringe exists. In turn, this requires the existence of the quantum operator phase. Therefore the state operator is coherent and, consequently, fundamentally non-hermitian.

## APPENDIX

It is shown by a short calculation that a microstate cannot survive the act of direct, invasive, measurement. An electron spin will interact with a magnetic field caused either by local moments or an external field. Assume that the external field is in the laboratory $\hat{\mathbf{Z}}$ direction, given by $\mathbf{H}_o = H_o \hat{\mathbf{Z}}$. Therefore the hamiltonian for the Zeeman interaction is,

$$H = -\boldsymbol{\mu}_e \cdot \hat{\mathbf{Z}} H_o \tag{A.1}$$

in terms of the magnetic moment of a spin, $\boldsymbol{\mu}_e$. Due to a spin having two quantization axes there is a magnetic moment associate with each. It is assumed that the Bohr magneton, $\beta$, and the g-factor are responsible for the magnetic moments along each quantization axis,

$$\boldsymbol{\mu}_{ez} = -\frac{1}{2} g \beta \sigma_z \hat{\mathbf{z}} \quad \text{and} \quad \boldsymbol{\mu}_{ex} = -\frac{1}{2} g \beta \sigma_x \hat{\mathbf{x}} \tag{A.2}$$

Due to the orthogonality of $\hat{\mathbf{z}}$ and $\hat{\mathbf{x}}$, the dipole-dipole interaction within a single electron is zero. The total magnetic moment is the difference between the two axes of quantization because, see Figure 1(a), the angular momentum polarization is directed oppositely along $\hat{\mathbf{z}}$ and $\hat{\mathbf{x}}$ in a microstate giving,

$$\boldsymbol{\mu}_e = \boldsymbol{\mu}_{ez} - \boldsymbol{\mu}_{ex} = -\frac{1}{2} g \beta \left( \sigma_z \hat{\mathbf{z}} - \sigma_x \hat{\mathbf{x}} \right) \tag{A.3}$$

The time dependence of a state using the Hamiltonian Eq.(A.1) is given by two separate equations,

$$\hat{\mathbf{z}}(t) = \hat{\mathbf{z}} \exp\left(+i\omega_o t \sin\theta\right) \quad \text{and} \quad \hat{\mathbf{x}}(t) = \hat{\mathbf{x}} \exp\left(-i\omega_o t \cos\theta\right) \tag{A.4}$$

with precessional frequency defined $\omega_o = g\beta H_o / 2$. The angle, $\theta$, is between the external magnetic field direction, $\hat{\mathbf{Z}}$ and the microstate vectors $\hat{\mathbf{z}}$ and $\hat{\mathbf{x}}$,

$$\hat{\mathbf{z}} \cdot \hat{\mathbf{Z}} = +\cos\theta \quad \text{and} \quad \hat{\mathbf{x}} \cdot \hat{\mathbf{Z}} = -\sin\theta \tag{A.5}$$

When $\theta = 0$ $\hat{\mathbf{z}}$ is aligned with the magnetic field direction, $\hat{\mathbf{Z}}$, so $\hat{\mathbf{z}}$ is a constant and $\hat{\mathbf{x}}$ presesses. Likewise for $\theta = \pi/2$ $\hat{\mathbf{x}}$ is aligned with $\hat{\mathbf{Z}}$ so $\hat{\mathbf{x}}$ is a constant and $\hat{\mathbf{z}}$ presesses.

This calculation is done within a single microstate and shows in the presence of a magnetic field that if one component aligns with the field, the other undergoes precessional motion in the plane perpendicular to the stationary axis. Over all the microstates with random phases, the precessional motion will average to zero leading to either one axis of quantization,

$$\overline{\hat{\mathbf{z}}(t)}^{3D} = \hat{\mathbf{Z}} \qquad \overline{\hat{\mathbf{x}}(t)}^{3D} = 0 \tag{A.6}$$

or the other,

$$\overline{\hat{\mathbf{z}}(t)}^{3D} = 0 \qquad \overline{\hat{\mathbf{x}}(t)}^{3D} = \hat{\mathbf{Z}} \tag{A.7}$$

thereby defining the macrostate, Eq.(3.6) with $\tilde{\mathbf{z}} = \hat{\mathbf{Z}}$ or with $\hat{\mathbf{x}} = \hat{\mathbf{Z}}$. Therefore preparation for measurement by an external probe removes one axis of quantization leaving only one element of physical reality for a spin. The other two components of physical reality are time averaged and phase randomized away in the interval that is practical for state preparation and measurement. Attempts to measure the spin in any other direction is therefore impossible unless the system is again prepared in that direction by re-orienting the external magnetic field. Figure 1(c) is thus the usual view of a pure spin state and corresponds to statistical ensembles of prepared states as discussed by Ballentine [19]. In particular in his statistical theory of measurement, he shows that coupling to an external apparatus takes time and during this time the microscopic physical state can change, consistent with the averages Eqs.(A.6) and (A.7).

## ACKNOWLEDGEMENTS

This work is supported by a Discovery Grant from the Natural Sciences and Engineering Research Council of Canada (NSERC).

## REFERENCES

1  B. C. Sanctuary, "Separation of Bell States", J. Phys. A. (submitted), and arXiv: quant-ph.
2  A. Einstein, B. Podolsky, and N. Rosen, "Can quantum-mechanical description of physical reality be considered complete?" Phys. Rev. 1935 **47**, 777.
3  G. Weihs, T. Jennewein, C. Simon, H. Weinfurter, A. Zeilinger, , "Violation of Bell's inequality under strict Einstein locality conditions", *Phys.Rev.Lett*. 1998 **81** 5039-5043 .
4  J. S. Bell, "On the Einstein-Podolsky-Rosen Paradox." Physics 1964 **1** 195; Bell, J. S.: *Speakable and Unspeakable in Quantum Mechanics*, Cambridge University Press; New York, Cambridge University Press, 1987.
5  B. C. Sanctuary, "Rationalization of EPR experiments.", In AIP Conference Proceedings, Växjö Conference: *Quantum Theory: Reconsideration of Foundations, 4.* Ed. A. Khrennikov, submitted, and arXiv: quant-ph..
6  A.Tonomura, J. Matsuda, T. Endo, T. Kawasaki, and H. Ezawa, , "Demonstration of Single-Electron buildup of an Interference Pattern", Am. J. Phys. 1989 **57**, 117.
7  K. Laidler, J. Meiser, and B.C. Sanctuary, *Physical Chemistry*, Houghton Mifflin, 4th edition, 2003.
8  R. C. Tolman, "The Principles of Statistical Mechanics." Oxford, London, 1938.
9  Here measurable means that the quantity has a handle, such as a magnetic moment, which in principle, if not fact, makes it measurable.
10  U. Fano, , "Description of States of Quantum Mechanics by Density Matrix and Operator Techniques." Rev. Mod. Phys. 1957 **29**, 74.
11  J. F. Clauser and A. Shimony, "Bell's Theorem: Experimental Tests and Implications", Rep. Prog. Phys., 1978, **41**, 1881–1927.
12  A. Aspect, "Bell's Theorem: the Naïve View of an Experimentalist" in *Quantum [Un]speakables – From Bell to Quantum information*., edited by R. A. Bertlmann and A. Zeilinger, Springer 2002, 1–34..
13  B. C. Sanctuary, , "Correlations in Entangled States" International J. of Physics B 2006 **20**, 1496.
14  J. F. Clauser, M.A. Horne, A. Shimony and R. A. Holt, "Proposed experiment to test local hidden-variable theories, Phys. Rev. Lett. 1969 **23**, 880-884.
15  K.Gustafson, "Noncommutative Trigonometry and Quantum Mechanics, in Advances in Deterministic and Stochastic Analysis." (N.Chuong, P.Ciarlet,P.Lax,D.Mumford, D.Phong, eds.), World Scientific 2007, p381. (Note that the angles used here, **a**,**b**,**b'**,**c** correspond to Gustafson's **d**,**b**,**a**,**c** .)
16  J. S. Bell, in *Foundations of Quantum Mechanics*, Proceedings of the International School of Physics "Enrico Fermi", Course XLIX, B. d'Espagnat (Ed.) (Academic, New York, 1971),
17  D. Bohm, *Quantum Theory*, Prentice-Hall 1951.
18  I. Marcikic, H. de Riedmatten, W. Tittel, H. Zbinden, N. Gisin, *Long-Distance Teleportation of Qubits at Telecommunication Wavelengths*, Nature, 2003 **421**, 509.
19  L. E. Ballentine, "The Statistical Interpretation of Quantum Mechanics", *Rev. Mod. Phys*., 1970, **42**, 358–381.
20  L. E. Ballentine, , "Quantum Mechanics, A Modern Development." World Scientific Publishing Co. Ltd., 2000.
21  M .Kupczynski, "Seventy Years of the EPR Paradox", in Albert Einstein Century Conference, AIP Conference Proceedings, Volume 861, ed. Jean-Michel Alimi and Adré Füzfa, New York, AIP Melville, 2006, pp.516-523
22  J. A. Khrennikov, *Interpretation of Probabilities*, Utrecht, VSP 1999.

23 E. Schrödinger, "Discussion of Probability Relations Between Separated Systems.", Proceedings of the Cambridge Philosophical Society, 1935 **31**: 555-563.
24 I. Pitowsky, "Resolution of the Einstein-Podolsky-Rosen and Bell Paradoxes.", Phys. Rev. Lett. 1982, **48**, 1299-1302.
25 M. Kupczynski, "Pitowsky Model and Complementarity." *Phys. Lett. A,* 1987, **121**, 51-53.
26 M. Kupczynski, "On the completeness of quantum mechanics.", arXiv: quant-ph/0208061, 9 August, 2002.
27 K. Hess and W. Philipp, Europhys. Lett., **57**:775 (2002)
28 J. F. Clauser and M. A. Horne, *Experimental consequences of objective local theories*, Physical Review D, **10**, 526-35 (1974)
29 L. Accardi and M. Regoli, "Locality and Bell's Inequities" in QP-XIII, *Foundations of Probability and Physics*, ed. A. Khrennikov, Singapore, World Scientivic, 2002, 1-28.
30 Bohr, N., "Can quantum-mechanical description of physical reality be considered complete?" Phys. Rev. **48**, 696 (1935).